\documentclass[aps, floatfix, reprint, longbibliography, 10pt]{revtex4-1}

\pdfoutput=1
\usepackage{graphicx}
\usepackage[T2A]{fontenc}
\usepackage[utf8]{inputenc}
\usepackage[english]{babel}
\usepackage{amsmath}
\usepackage{hyperref}
\hypersetup{
  colorlinks   = true, 
  urlcolor     = blue, 
  linkcolor    = black, 
  citecolor    = blue  
}

\begin{document}

\title{Optical Limiting Performance of a GaAs/AlAs Heterostructure Microcavity\\ in the Near-Infrared}

\author{A.\,A.~Ryzhov}\email{a\_ryzhov@niuitmo.ru}
\affiliation{Institute for Laser Physics, Vavilov State Optical Institute, Kadetskaya lin. 5/2, St. Petersburg, 199034, Russia}
\affiliation{Laboratory of Nonlinear Optical Informatics, ITMO University, Kronverkskiy pr. 49, St. Petersburg, 197101, Russia}

\begin{abstract}
A multilayer GaAs/AlAs heterostructure forming a Fabry--Perot microcavity with a narrow resonance at 1.1~$\mu$m was produced by molecular-beam epitaxy. Under nanosecond pulsed laser radiation, a blue shift of the resonant line, associated  with a photo-induced negative change in refractive index in GaAs, was experimentally registered by using an optical parametric oscillator. The spectral shift was accompanied by a reduction in peak transmittance, associated with nonlinear intracavity absorption. Such a cavity can be used as an optical limiter at the resonant wavelength when both the spectral shift and the transmittance reduction contribute to the limiting effect. An exceptionally low limiting threshold of about 1 mJ/cm$^2$ was observed in the experiment.
\end{abstract}

\maketitle

\section{Introduction}

Nonlinear Fabry--Perot resonators have been the subject of great interest since the 1970s as promising optical computing elements \cite{Lugovoi1979, Abraham1982}. Whereas initially large conventional resonators, in which the distance between the two mirrors far exceeds the wavelength, were under consideration, solid optical microcavities now seem to be quite applicable in optical systems and optoelectronics~\cite{Vahala2003}. Microcavity mirrors are typically formed by alternating layers of dielectric or semiconducting media with different refractive indices so that a microcavity in whole represents a one-dimensional photonic crystal with a defect. The first observation of optical bistability in a thin-film multilayer resonator was reported at least as early as in 1978~\cite{Karpushko1978}. Semiconductor microcavities are of particular interest owing to a variety of strong optical nonlinearities and very high structural perfection provided by molecular-beam epitaxy~\cite{Deveaud2007}.

Besides optical data processing, the main purpose of optical limiting is to protect optical sensors from excessively powerful laser pulses that may cause damage. There have been many different approaches to developing optical limiters since the 1970s~\cite{Tutt1993, Hollins1999}. Nevertheless, there is still a lack of optical limiters applicable to protection of optical sensors. The problem is that the principles of optical limiters are based on various nonlinear optical effects, all of which require relatively high light intensity. As a result, in the majority of cases a limiter starts to reduce its transmittance at a level of incident intensity that is much higher than that needed for protection of the optical sensor. Consequently, one of the main tasks in development of optical limiters is to reduce the optical limiting threshold. This is a challenging task given that an optical limiter should not significantly reduce desired signals so that the initial transmission should be as close to 100\% as possible. 

The present paper is devoted to the limiting of single nanosecond pulses, which are typical for Q-switched lasers. As far as 1992 the limiting threshold of ${\sim0.1}$~J/cm$^2$ was reported for fullerene solutions, using 8-ns pulses at 532 nm~\cite{Tutt1992}. Today's representative values of limiting threshold in various composite materials providing nonlinear absorption or nonlinear scattering are still in the range 0.1--1~J/cm$^2$ under nanosecond pulses for visible and near-infrared radiation~\cite{Belousova2014, Shanshool2016, Mikheev2016}. Moreover, most studies deal with solutions, while for practical applications rather solid forms are desirable.

A leap forward was made with graphene-polymer films, in which the optical limiting behaviour is associated with a novel reverse saturable absorption mechanism. The threshold of $\sim$0.01~J/cm$^2$ under 3.5-ns pulses at both 532 and 1064~nm for a 2-$\mu$m-thick film was reported in 2011, providing a new benchmark for the broadband optical limiting~\cite{Lim2011}. A comparable result, the threshold of $\sim$0.01~J/cm$^2$ under 16~ns pulses at 532~nm, has been obtained with a solution of a phthalocyanine J-type dimeric complex of Zn very recently~\cite{Tolbin2016}.

Many works involve ultrashort pulses where thresholds in terms of fluence are much lower but in terms of peak intensity are much higher. Attempting to extrapolate results obtained with pico- or femtosecond lasers to the nanosecond field would be inappropriate because usually different nonlinear mechanisms are involved at different pulse durations.

A nonlinear Fabry--Perot cavity acts as an optical limiter at its resonant wavelength when the intracavity medium changes its refractive index or increases its absorption under irradiation. A nonlinear microcavity can have a much lower limiting threshold compared with a bulk material for two reasons. Firstly, at the resonant wavelength, the electromagnetic field is localized within the cavity so that the intracavity intensity is much higher than the external intensity. Secondly, the cavity is a narrow-band optical filter and its resonant line of transmittance is highly sensitive to any change in the intracavity medium optical parameters. A small change in refractive index leads to a significant spectral shift of the line while a small additional absorption within the cavity leads to a significant peak transmittance reduction. The optical limiting characteristics of Fabry–Perot microcavities in the simplest case of third-order nonlinear absorption and refraction of the intracavity medium have been considered in detail in~\cite{Ryzhov2015}.

Of course, there is a principal restriction connected with the fact that a resonator has a narrow resonant line of transmission. This means that a resonator can be used as a limiter only at this particular wavelength while all radiation in the vicinity of this wavelength is reflected. However, there does exist a class of active laser systems as optical detection and ranging systems to which such a limiter can be applicable. It is also noteworthy that in the earliest proposal of an optical power limiter in 1962 a nonlinear Fabry--Perot resonator was considered as an ideal reflective limiter\cite{Siegman1962}.

There are a few recent works experimentally demonstrating optical limiting performance of thin-film resonators~\cite{Ryzhov2014, Valligatla2015, Dong2015, Vella2016}. In~\cite{Ryzhov2014} and \cite{Valligatla2015}, microcavities were made of dielectric layers by vacuum evaporation with $\mathrm{Nb_2O_5}$ and ZnO, respectively, as nonlinear intracavity media. In both works, the limiting characteristics were measured at 532~nm under 5--7 ns pulses. Different results were obtained though: in \cite{Ryzhov2014}, a limiting threshold of ${\sim5\times 10^{-3}}$~J/cm$^2$ with 75\% of initial transmission was observed, while the authors of~\cite{Valligatla2015} claim a threshold 0.74~J/cm$^2$ with only 6\% of initial transmission. In~\cite{Ryzhov2014} it was also shown experimentally that the decrease in transmittance accompanies a respective increase in reflectance.

In~\cite{Vella2016}, a microcavity with an amorphous GaAs layer placed between its dielectric mirrors was fabricated by plasma-enhanced chemical vapour deposition. The authors got a resonant line at ${\sim1.6}$~$\mu$m with 50\% of initial transmission. Usage of GaAs as a nonlinear intracavity medium for near-infrared radiation makes this work seem similar to the presented paper; however, in~\cite{Vella2016}, the nonlinear characteristics of the cavity were studied under a completely different laser irradiation (150-fs pulses at 1~kHz). As well as in the present paper, in~\cite{Vella2016}, transmittance spectra under low-intensity and high-intensity radiation were measured, and a slight broadening of the transmittance line was observed with no spectral shift. At the resonant wavelength, a great reduction in transmittance accompanied by a corresponding growth in reflectance was observed. The authors evaluated the limiting threshold as 25~GW/cm$^2$ in terms of peak intensity. In terms of the single pulse fluence, this is ${\sim3\times 10^{-3}}$~J/cm$^2$, which is of the same order of magnitude as the value reported in the present paper.

\section{Microcavity structure}

\begin{figure}
\centering
\includegraphics[width=0.6\columnwidth]{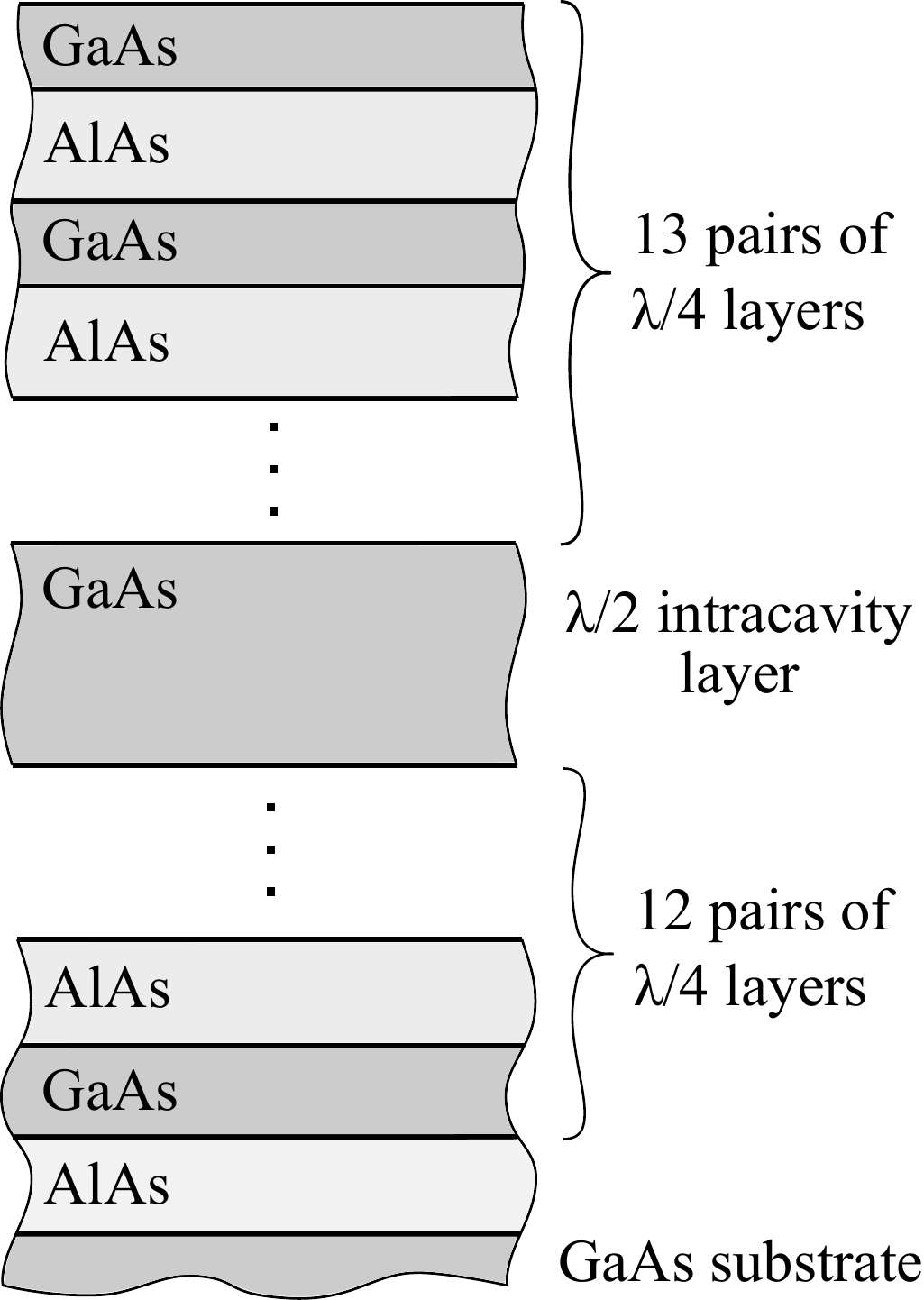} 
\caption{Scheme of the GaAs/AlAs heterostructure microcavity.}
\label{fig:structure}
\end{figure}

The multilayer GaAs/AlAs heterostructure was grown by molecular-beam epitaxy on a GaAs substrate using the Riber~21 system. The whole structure consists of 52 alternating layers, as shown in~Fig.~\ref{fig:structure}, and forms a Fabry--Perot microcavity with a resonance at $\lambda_0 = 1116.8$~nm. The value of $\lambda_0$ is double optical thickness of the middle GaAs layer. The refractive indices of GaAs and AlAs at 1.1 $\mu$m at room temperature are 3.49 and 2.95, respectively~\cite{Adachi1985, Deri1995}. This means that the geometrical thicknesses of the $\lambda/4$ layers are 76.2~nm (GaAs) and 90.1~nm (AlAs) while the intracavity layer of GaAs is 152.4~nm.

\begin{figure}
\centering
\includegraphics[width=\columnwidth]{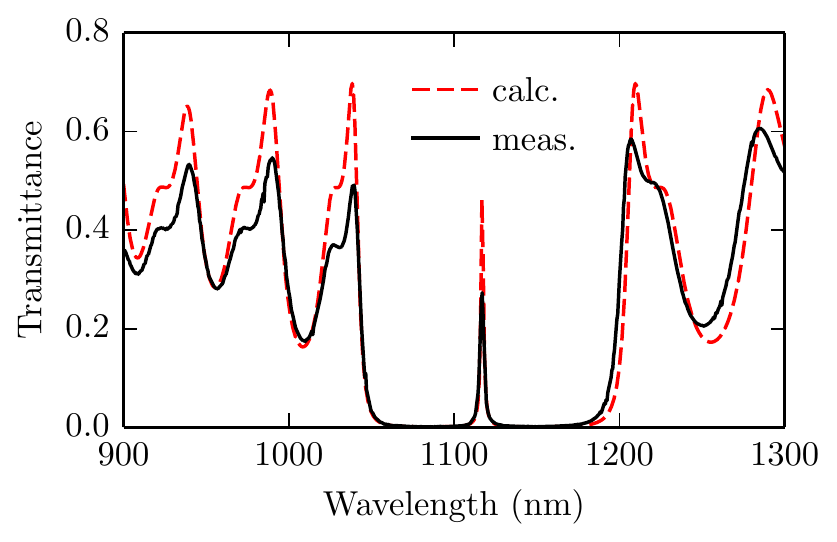} 
\caption{Measured and calculated transmission spectra of the GaAs/AlAs heterostructure.}
\label{fig:spectrum}
\end{figure}

The transmittance spectrum of the heterostructure was measured by Shimadzu UV-3600 spectrophotometer and is presented in Fig.~\ref{fig:spectrum}. The initial transmission at the resonant wavelength is about half of what it was expected by calculation, presumably owing to some inaccuracy in polishing of the substrate. It should be noted that the backside of the substrate was not bloomed and 30\% incoherent reflectance at the GaAs--Air interface plays a role. A good agreement in extrema positions between the measured and the calculated spectra verifies the layers' thicknesses though.

The upper mirror (bordering with air) of the sample includes 13 layer periods and has 98.6\% reflectance. The lower mirror consists of 12.5 layer periods as the last GaAs layer would be inessential on a GaAs substrate. The reflectance of the lower mirror is 95\% (incoherent reflection on the substrate--air interface is set aside). Generally, more layers from the substrate side would give a transmittance value close to that from the air side and consequently almost 100\% of transmittance at $\lambda_0$. The substrate--air interface needs blooming though to avoid transmittance reduction through the incoherent reflection.

\begin{figure}
\centering
\includegraphics[width=\columnwidth]{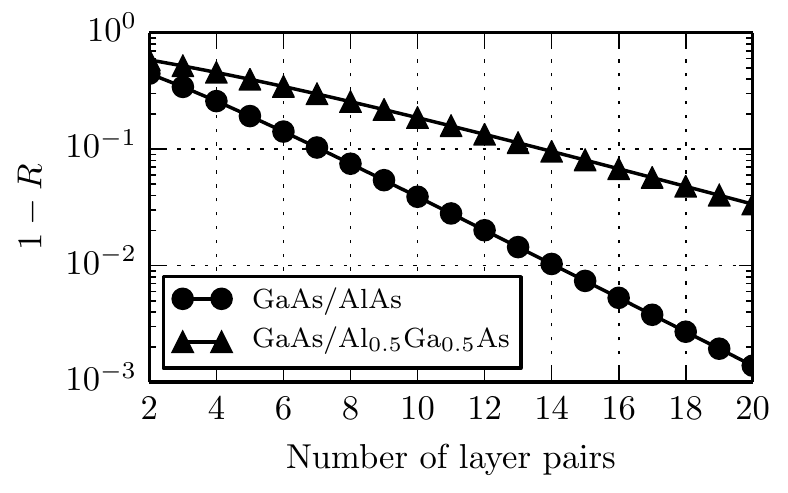} 
\caption{Dependence of the Bragg mirror reflectance on the number of layer pairs for GaAs/Al$_x$Ga$_{1-x}$As at $x=1$ and $x=0.5$. For better readability of reflectance close to unity, $(1-R)$ instead of $R$ is used as the ordinate.}
\label{fig:mirror}
\end{figure}

Figure~\ref{fig:mirror} shows the dependence of the GaAs/AlAs multilayer mirror reflectance on the number of $\lambda/4$ layers (where half-infinite air from the GaAs side and half-infinite GaAs from the AlAs side are assumed). To illustrate the importance of the refractive index contrast between the layers, the same dependence is shown for GaAs/Al$_{0.5}$Ga$_{0.5}$As mirrors. The refractive index of Al$_{0.5}$Ga$_{0.5}$As is 3.23, which is exactly midway between that of AlAs and GaAs~\cite{Adachi1985, Deri1995}. For example, a 28-layer (i.e., consisting of 14 pairs) GaAs/AlAs mirror has a reflectance of 99\% whereas the same number of layers of GaAs/Al$_{0.5}$Ga$_{0.5}$As gives only 90\% reflectance.

\begin{figure}
\centering
\includegraphics[width=\columnwidth]{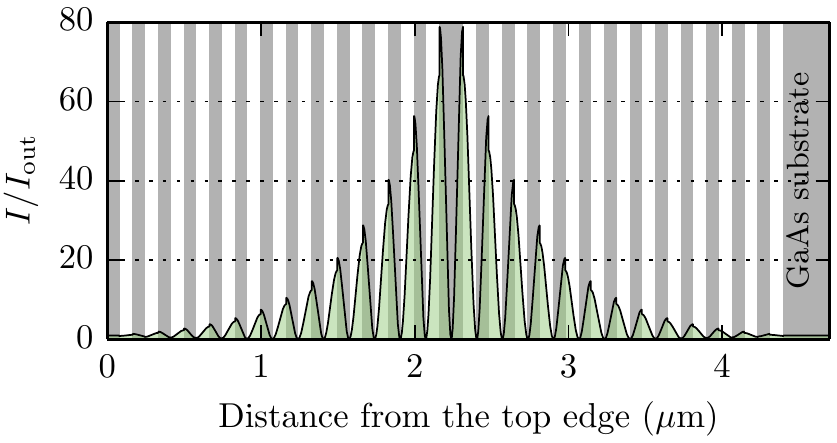} 
\caption{Light intensity distribution as the ratio of local intensity $I$ to output intensity $I_\mathrm{out}$ within the 52-layer GaAs/AlAs microcavity at the resonant wavelength. The grey colour indicates GaAs layers, the white --- AlAs layers.}
\label{fig:field}
\end{figure}

Figure~\ref{fig:field} shows the calculated intensity distribution within the sample at $\lambda_0$ when no nonlinearity is in action. Here the layers are depicted as vertical. The heterostructure is illuminated normally from the air side and the substrate is assumed to be semi-infinite. The calculated transmittance at $\lambda_0$ is 69\%. As shown in the figure, the intensity in the loops of the standing wave within the cavity is a factor of 80 higher than the output intensity. This corresponds to 95\% reflectance of the output mirror since the peak intracavity intensity $I_\mathrm{max}$ obeys the relation
\begin{equation}
I_\mathrm{max} = \frac{4 I_\mathrm{out}}{1-R_\mathrm{out}},
\end{equation}
where $I_\mathrm{out}$ is the output intensity and $R_\mathrm{out}$ is the reflectance of the output mirror.

The GaAs and AlAs energy gaps are 1.42 and 2.16~eV, respectively, while the photon energy at $\lambda_0 = 1117$~nm is 1.11~eV. Therefore, both materials are transparent at $\lambda_0$ but two-photon absorption (2PA) in GaAs can be sufficient to make the microcavity optically nonlinear for short pulses. 2PA in GaAs at ${\sim1.1}$~$\mu$m has been studied in many works; it can be assumed with confidence that 2PA coefficient ${\beta = 25}$~cm/GW \cite{Said1992}.

\section{Experimental results and discussion}

\begin{figure}
\centering
\includegraphics[width=\columnwidth]{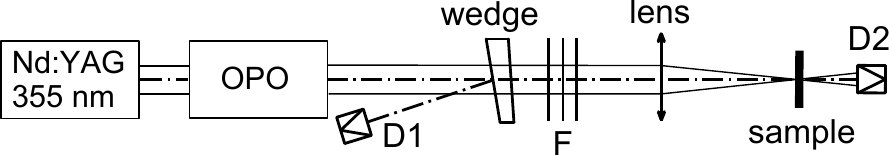} 
\caption{Scheme of the nonlinear transmission measurements. OPO: optical parametric oscillator; D1, D2: laser pulse energy meters; F: set of grey glasses.}
\label{fig:setup}
\end{figure}

By using an optical parametric oscillator (OPO), changes of the transparency line of the microcavity under pulsed laser radiation were registered. The scheme of the experiment is presented in Fig.~\ref{fig:setup}. The OPO was pumped by 10-ns pulses at 355~nm and the sample was irradiated by single pulses of the same duration in a small vicinity of the resonant wavelength. The beam was focused by a 60~mm focal length lens on the sample so that the area encircling 86\% of the pulse energy was ${\sim10^{-3}}$~cm$^2$. This estimation results from the fact that 75\% of the pulse energy went through a 0.3-mm aperture and the assumption that the focal spot was Gaussian. The energy of incident pulses was varied over a wide range by a set of absorbing glasses. The absorbing glasses were placed before the focusing lens where the laser beam was 8-mm wide, therefore no nonlinear effect was possible within them.

\begin{figure*}
\centering
\includegraphics[width=\textwidth]{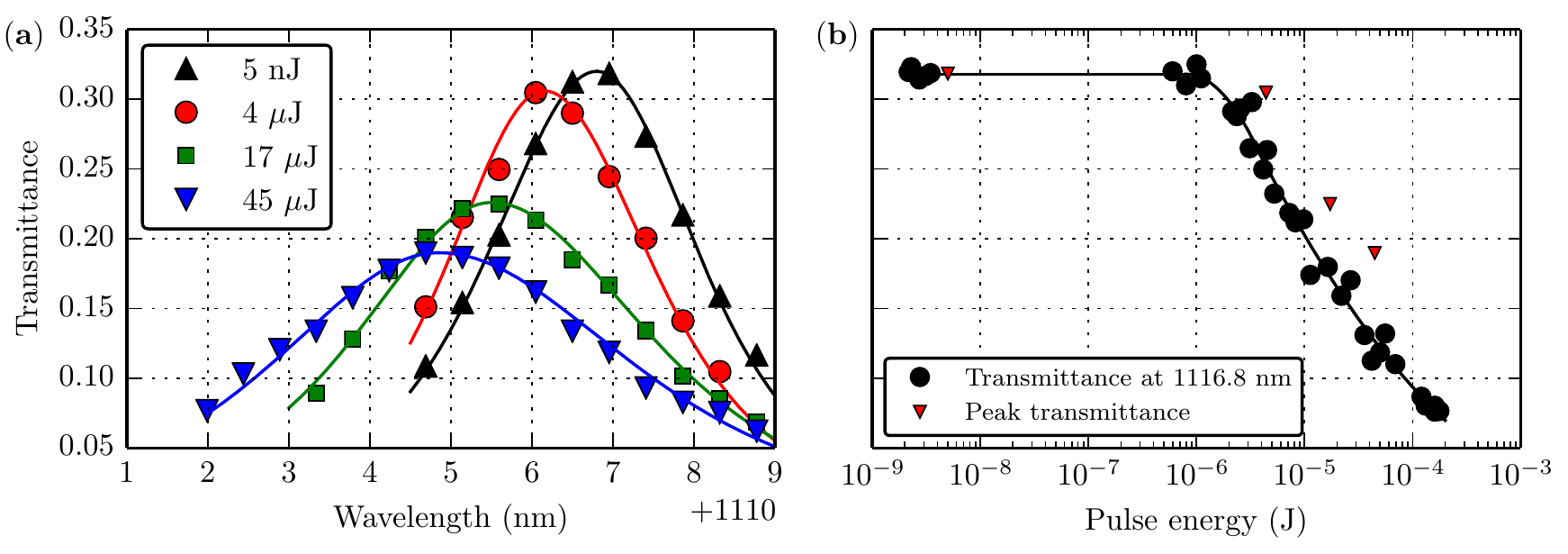}
\caption{Nonlinear transmittance behaviour of the GaAs/AlAs microcavity under pulsed laser radiation. The laser spot area is ${\sim10^{-3}}$~cm$^2$. (a) Resonant line of transmission at different values of incident pulse energy. (b) Optical limiting at the resonant wavelength (black circles) along with the dependence of peak transmittance on pulse energy (red triangles). The peak transmittance dependence was retrieved from the same data that were used for plotting (a), while the optical limiting characteristic was obtained from a separate measurement performed at the fixed wavelength $\lambda_0 = 1116.8$~nm.}
\label{fig:nonlinear_response} 
\end{figure*}

In the course of the experiment, at a fixed level of incident pulse energy the wavelength was varied by OPO in a small vicinity of the resonant wavelength. Obtained contours of the resonant transmission line at four different values of pulse energy are presented in Fig.~\ref{fig:nonlinear_response}a. The plotted points correspond directly to the experimental data while the solid lines are their approximations by the Lorentz function
\begin{equation}
T = T_0 \left(1 + \left(\frac{\lambda - \lambda_0}{w}\right)^2 \right)^{-2},
\label{eq:Lorentz}
\end{equation}
where the width parameter $w$ is different for the left and right sides ($\lambda < \lambda_0$ and $\lambda > \lambda_0$ areas); i.e., the contours are asymmetric. The parameters of the approximating contours are presented in Table~\ref{tab:shift}. It can be summarized that with increasing pulse energy the resonant line gets lower, wider and more asymmetric. 

\begin{table}
\caption{Approximation parameters for Fig.~\ref{fig:nonlinear_response}a in accordance with~\eqref{eq:Lorentz}. $w_\mathrm{blue}$ and $w_\mathrm{red}$ relate to the areas $\lambda < \lambda_0$ and $\lambda > \lambda_0$ correspondingly.}
\centering
\begin{tabular}{ccccc}
\hline 
$E_\mathrm{pulse}$ ($\mu$J) & $T_0$ & $\lambda_0$ (nm) & 
$w_\mathrm{blue}$ (nm) & $w_\mathrm{red}$ (nm) \\ 
\hline 
0.005 & 0.32 & 1116.8 & 2.45 & 2.3 \\ 
4 & 0.306 & 1116.15 & 2.2 & 2.45 \\ 
17 & 0.226 & 1115.5 & 3.0 & 3.5 \\ 
45 & 0.19 & 1114.85 & 3.7 & 4.3 \\ 
\hline 
\end{tabular} 
\label{tab:shift}
\end{table}

In relation to the limiting effect, Fig.~\ref{fig:nonlinear_response}a shows that at 4~$\mu$J the decrease in peak transmittance was small (from 32\% to 30.6\%). At the same time, the spectral shift of ${\sim0.65}$~nm decreased the transmittance at the initially resonant wavelength ($\lambda_0 = 1116.8$~nm) to 27\%. 
At higher pulse energies (17 and 45 $\mu$J), the transmittance reduction at $\lambda_0$ is also mainly associated with the spectral shift rather than with the peak transmittance reduction. 

In the next experiment with the same setup, the optical limiting characteristic at $\lambda_0$ was obtained (Fig.~\ref{fig:nonlinear_response}b). The limiting threshold, at which the transmittance starts to go down, can be estimated at $10^{-3}$ J/cm$^2$. For comparison, associated with 2PA and consequent free-carrier absorption optical limiting in bulk InP (InP has similar to GaAs band gap, 2PA-coefficient and free-carrier absorption cross-section) would give the threshold of $\sim0.1$ J/cm$^2$ under 10-ns pulses at 1.1~$\mu$m~\cite{Gonzalez2009}. The dependence of peak transmittance on pulse energy, retrieved from the experiment with varying wavelength (Fig.~\ref{fig:nonlinear_response}a), is also shown in Fig.~\ref{fig:nonlinear_response}b to additionally illustrate relative contribution of the spectral shift to the decrease in transmittance at $\lambda_0$.

While measuring, the sample was not moved relative to the laser spot, i.e., always the same area of the sample was irradiated. However, the laser pulses were single (button-activated) and went with intervals of at least 1~s. Under these conditions, no cumulative effect was expected and observed: the transmittance fell with increasing pulse energy and rose back to its initial value with decreasing pulse energy. Furthermore, no optical damage occurred within the energy range presented in Fig.~\ref{fig:nonlinear_response}b. A higher energy would cause a breakdown, visible on the surface, which was previously determined in another area of the sample.

Generally, peak transmittance reduction results from nonlinear growth of intracavity absorption. Under nanosecond pulses, both 2PA itself and additional absorption of 2PA-generated free carriers play a role~\cite{Gonzalez2009, Boggess1985}. The blue spectral shift results from a negative change in the real refractive index of the intracavity layer. This change is proportional to the spatial density of photo-induced free carriers with the coefficient $\sigma_\mathrm{r} \simeq 7\cdot10^{-21}$~cm$^3$~\cite{Said1992}. Free carriers can be generated as a result of either 2PA or residual linear absorption by impurities in GaAs.

The rate of free carrier generation in the presence of linear absorption and 2PA is defined by the equation
\begin{equation}
\frac{dN}{dt} = \frac{I}{h\nu} \left( \alpha + \frac{\beta I}{2} \right),
\label{eq:fc_rate}
\end{equation}
where $N$ is free carrier concentration, $I$ is field intensity, $\alpha$ is linear index of absorption and $\beta$ is 2PA-coefficient. For GaAs at 1.1~$\mu$m ${\beta = 25}$~cm/GW \cite{Said1992}. 

The peak intensity within the cavity at the limiting threshold can be roughly calculated as
\begin{equation}
I_\mathrm{peak} = 80 \frac{10^{-3}\mathrm{\ J/cm^2}}{10^{-8} s} = 8 \cdot 10^6\mathrm{\ W/cm^2},
\end{equation}
where the factor 80 is the field enhancement factor taken from Fig.~\ref{fig:field}, $10^{-3}\mathrm{\ J/cm^2}$ is the limiting threshold value and $10^{-8}$~s is the pulse duration. Thus, the second term in~\eqref{eq:fc_rate} can be estimated as 
\begin{equation}
\frac{\beta I}{2} \simeq 0.1 \mathrm{\ cm^{-1}}.
\label{eq:betaI/2}
\end{equation}

As it was shown by the present experiment, under nanosecond pulses optical limiting in the cavity with pure GaAs is mainly related to the refractive index change rather than to an increase in absorption. This suggests that moderate intracavity linear absorption could sufficiently decrease the limiting threshold. Such absorption should be low enough to not significantly reduce the initial resonant transmission of the cavity. 

At the resonant wavelength, linear transmittance of a Fabry--Perot cavity in the presence of intracavity linear absorption
\begin{equation}
T = \frac{(1 - R_1) (1 - R_2)}{(1 - T_1 \sqrt{R_1 R_2})}T_1,
\end{equation}
where $R_1, R_2$ are reflectances of the mirrors, $T_1$ is single-pass transmittance of the intracavity layer. The transmittance reduction related to such linear absorption can be expressed as
\begin{equation}
\frac{T}{T_0} = T_1 \left( \frac{1 - \sqrt{R_1 R_2}}{1 - T_1 \sqrt{R_1 R_2}} \right)^2,
\label{eq:T2T0}
\end{equation}
where $T_0$ is the cavity transmittance at $T_1 = 1$.

\begin{figure}
\centering
\includegraphics[width=\columnwidth]{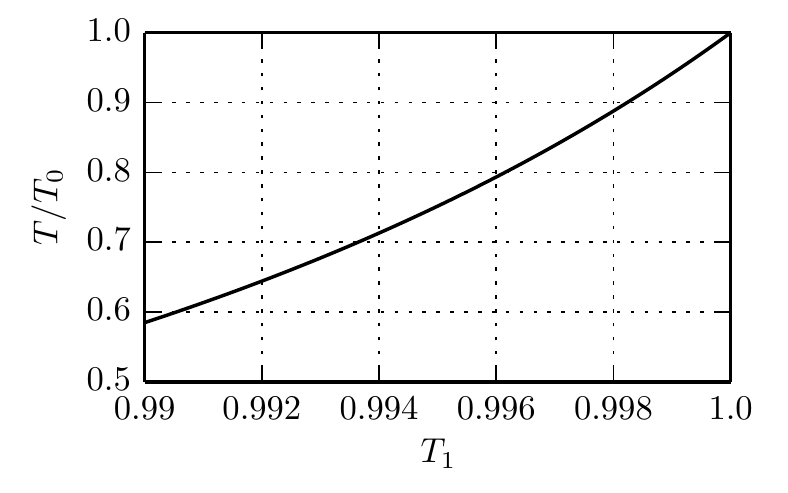}
\caption{Calculated relative transmittance of the GaAs/AlAs heterostructure as a function of single-pass transmittance $T_1$ of the intracavity layer.}
\label{fig:T2T0} 
\end{figure}

For the given heterostructure $R_1 = 0.986, R_2 = 0.950$. The dependence of $T/T_0$ on $T_1$ in accordance with~\eqref{eq:T2T0} is plotted in~Fig.~\ref{fig:T2T0}.

As one can see from the plot, at $T_1 = 0.998$ relative transmittance reduction is about 10\% ($T \simeq 0.9 T_0$) which is probably acceptable for practical purposes. The linear absorption index, corresponding to this value of $T_1$, is
\begin{equation}
\alpha = \frac{(1 - T_1)}{d} = 131 \mathrm{\ cm^{-1}}, 
\end{equation}
where $d = 152.4$~nm is the intracavity layer thickness.

This value of $\alpha$ is four orders of magnitude higher than the value of $\beta I / 2$ (\eqref{eq:betaI/2}), which means, according to~\eqref{eq:fc_rate}, that such intracavity absorption would give a four orders of magnitude higher rate of free carrier generation at the given value of $I_\mathrm{peak}$. Or, on the other hand, it would give the same rate at a four orders of magnitude lower value of $I_\mathrm{peak}$, which would lead to a proportional reduction of the limiting threshold. Such a huge reduction of the limiting threshold at the expense of several percent of linear absorption loss shows great promise. Presumably, a desired amount of absorption could be achieved by adding some impurities or implementing a narrow but highly absorbing quantum well within the intracavity semiconductor layer. This question requires a further detailed examination.

It was shown theoretically in~\cite{Andrews2016} that even a cavity with a saturable absorber may exhibit slight optical limiting owing to a refractive index change and a corresponding spectral shift.

\section{Conclusion}

A 52-layer GaAs/AlAs heterostructure, forming a Fabry--Perot microcavity with a resonance at 1.1~$\mu$m, was grown on a GaAs substrate by molecular-beam epitaxy. Good agreement in extrema positions between the measured and the calculated spectra was obtained, which verifies the composition (Fig.~\ref{fig:spectrum}).

By using an optical parametric oscillator, a blue spectral shift of the transparency line under pulsed laser radiation was clearly observed (Fig.~\ref{fig:nonlinear_response}a). An important finding is that, under nanosecond pulses, optical limiting in the GaAs/AlAs microcavity is mainly related to the intracavity refractive index change rather than to intracavity nonlinear absorption itself. At the same time, the refractive index change is induced by the nonlinear absorption.

An exceptionally low limiting threshold (for nanosecond pulses) of ${\sim10^{-3}}$~J/cm$^2$ was registered (Fig.~\ref{fig:nonlinear_response}b). However, the achieved transmittance reduction was only about 4 times when the incident pulse energy was upper-bounded by radiation resistance of the sample. 

Given the key role of photo-induced change in refractive index, it was suggested that implementing linear intracavity absorption for the sake of efficient free-carrier generation would sufficiently decrease the limiting threshold. As a first approximation, for the given heterostructure it was estimated that linear intracavity absorption at a level of $10^2$~cm$^{-1}$ would decrease the limiting threshold by four orders of magnitude at the expense of relative linear transmittance reduction by $\sim10\%$. This opens up promising opportunities for further research.

\section*{Acknowledgements}

The author thanks A.I.~Khrebtov, R.R.~Reznik, Prof. G.E.~Cirlin from St.~Petersburg Academic University for the growth of the heterostructure and Prof. I.M.~Belousova from Vavilov State Optical Institute for her encouragement of the research. 

The work was financially supported by the Government of the Russian Federation (074-U01) and the Russian Foundation of Basic Research (14-02-00851 A).

%


\begin{thebibliography}{25}%
\makeatletter
\providecommand \@ifxundefined [1]{%
 \@ifx{#1\undefined}
}%
\providecommand \@ifnum [1]{%
 \ifnum #1\expandafter \@firstoftwo
 \else \expandafter \@secondoftwo
 \fi
}%
\providecommand \@ifx [1]{%
 \ifx #1\expandafter \@firstoftwo
 \else \expandafter \@secondoftwo
 \fi
}%
\providecommand \natexlab [1]{#1}%
\providecommand \enquote  [1]{``#1''}%
\providecommand \bibnamefont  [1]{#1}%
\providecommand \bibfnamefont [1]{#1}%
\providecommand \citenamefont [1]{#1}%
\providecommand \href@noop [0]{\@secondoftwo}%
\providecommand \href [0]{\begingroup \@sanitize@url \@href}%
\providecommand \@href[1]{\@@startlink{#1}\@@href}%
\providecommand \@@href[1]{\endgroup#1\@@endlink}%
\providecommand \@sanitize@url [0]{\catcode `\\12\catcode `\$12\catcode
  `\&12\catcode `\#12\catcode `\^12\catcode `\_12\catcode `\%12\relax}%
\providecommand \@@startlink[1]{}%
\providecommand \@@endlink[0]{}%
\providecommand \url  [0]{\begingroup\@sanitize@url \@url }%
\providecommand \@url [1]{\endgroup\@href {#1}{\urlprefix }}%
\providecommand \urlprefix  [0]{URL }%
\providecommand \Eprint [0]{\href }%
\providecommand \doibase [0]{http://dx.doi.org/}%
\providecommand \selectlanguage [0]{\@gobble}%
\providecommand \bibinfo  [0]{\@secondoftwo}%
\providecommand \bibfield  [0]{\@secondoftwo}%
\providecommand \translation [1]{[#1]}%
\providecommand \BibitemOpen [0]{}%
\providecommand \bibitemStop [0]{}%
\providecommand \bibitemNoStop [0]{.\EOS\space}%
\providecommand \EOS [0]{\spacefactor3000\relax}%
\providecommand \BibitemShut  [1]{\csname bibitem#1\endcsname}%
\let\auto@bib@innerbib\@empty
\bibitem [{\citenamefont {Lugovoi}(1979)}]{Lugovoi1979}%
  \BibitemOpen
  \bibfield  {author} {\bibinfo {author} {\bibfnamefont {V.~N.}\ \bibnamefont
  {Lugovoi}},\ }\bibfield  {title} {\enquote {\bibinfo {title} {{Nonlinear
  optical resonators (excited by external radiation) (review)}},}\ }\href
  {\doibase 10.1070/QE1979v009n10ABEH009501} {\bibfield  {journal} {\bibinfo
  {journal} {Sov. J. Quantum Electron.}\ }\textbf {\bibinfo {volume} {9}},\
  \bibinfo {pages} {1207--1220} (\bibinfo {year} {1979})}\BibitemShut {NoStop}%
\bibitem [{\citenamefont {Abraham}\ and\ \citenamefont
  {Smith}(1982)}]{Abraham1982}%
  \BibitemOpen
  \bibfield  {author} {\bibinfo {author} {\bibfnamefont {E.}~\bibnamefont
  {Abraham}}\ and\ \bibinfo {author} {\bibfnamefont {S.~D.}\ \bibnamefont
  {Smith}},\ }\bibfield  {title} {\enquote {\bibinfo {title} {{Nonlinear
  Fabry-Perot interferometers}},}\ }\href {\doibase 10.1088/0022-3735/15/1/004}
  {\bibfield  {journal} {\bibinfo  {journal} {J. Phys. E.}\ }\textbf {\bibinfo
  {volume} {15}},\ \bibinfo {pages} {33--39} (\bibinfo {year}
  {1982})}\BibitemShut {NoStop}%
\bibitem [{\citenamefont {Vahala}(2003)}]{Vahala2003}%
  \BibitemOpen
  \bibfield  {author} {\bibinfo {author} {\bibfnamefont {K.~J.}\ \bibnamefont
  {Vahala}},\ }\bibfield  {title} {\enquote {\bibinfo {title} {{Optical
  microcavities}},}\ }\href {\doibase 10.1038/nature01939} {\bibfield
  {journal} {\bibinfo  {journal} {Nature}\ }\textbf {\bibinfo {volume} {424}},\
  \bibinfo {pages} {839--846} (\bibinfo {year} {2003})}\BibitemShut {NoStop}%
\bibitem [{\citenamefont {Karpushko}\ and\ \citenamefont
  {Sinitsyn}(1978)}]{Karpushko1978}%
  \BibitemOpen
  \bibfield  {author} {\bibinfo {author} {\bibfnamefont {F.~V.}\ \bibnamefont
  {Karpushko}}\ and\ \bibinfo {author} {\bibfnamefont {G.~V.}\ \bibnamefont
  {Sinitsyn}},\ }\bibfield  {title} {\enquote {\bibinfo {title} {{An optical
  logic element for integrated optics in a nonlinear semiconductor
  interferometer}},}\ }\href {\doibase 10.1007/BF00604756} {\bibfield
  {journal} {\bibinfo  {journal} {J. Appl. Spectrosc.}\ }\textbf {\bibinfo
  {volume} {29}},\ \bibinfo {pages} {1323--1326} (\bibinfo {year}
  {1978})}\BibitemShut {NoStop}%
\bibitem [{\citenamefont {Deveaud}(2007)}]{Deveaud2007}%
  \BibitemOpen
  \bibinfo {editor} {\bibfnamefont {B.}\ \bibnamefont {Deveaud}},\ ed.,\
  \href {\doibase 10.1002/9783527610150} {\emph {\bibinfo {title} {{The Physics
  of Semiconductor Microcavities}}}}\ (\bibinfo  {publisher} {Wiley-VCH},\
  \bibinfo {address} {Weinheim, Germany},\ \bibinfo {year} {2007})\BibitemShut
  {NoStop}%
\bibitem [{\citenamefont {Tutt}\ and\ \citenamefont
  {Boggess}(1993)}]{Tutt1993}%
  \BibitemOpen
  \bibfield  {author} {\bibinfo {author} {\bibfnamefont {L.~W.}\ \bibnamefont
  {Tutt}}\ and\ \bibinfo {author} {\bibfnamefont {T.~F.}\ \bibnamefont
  {Boggess}},\ }\bibfield  {title} {\enquote {\bibinfo {title} {{A review of
  optical limiting mechanisms and devices using organics, fullerenes,
  semiconductors and other materials}},}\ }\href {\doibase
  10.1016/0079-6727(93)90004-S} {\bibfield  {journal} {\bibinfo  {journal}
  {Prog. Quantum Electron.}\ }\textbf {\bibinfo {volume} {17}},\ \bibinfo
  {pages} {299--338} (\bibinfo {year} {1993})}\BibitemShut {NoStop}%
\bibitem [{\citenamefont {Hollins}(1999)}]{Hollins1999}%
  \BibitemOpen
  \bibfield  {author} {\bibinfo {author} {\bibfnamefont {R.~C.}\
  \bibnamefont {Hollins}},\ }\bibfield  {title} {\enquote {\bibinfo {title}
  {{Materials for optical limiters}},}\ }\href {\doibase
  10.1016/S1359-0286(99)00009-1} {\bibfield  {journal} {\bibinfo  {journal}
  {Curr. Opin. Solid State Mater. Sci.}\ }\textbf {\bibinfo {volume} {4}},\
  \bibinfo {pages} {189--196} (\bibinfo {year} {1999})}\BibitemShut {NoStop}%
\bibitem [{\citenamefont {Tutt}\ and\ \citenamefont {Kost}(1992)}]{Tutt1992}%
  \BibitemOpen
  \bibfield  {author} {\bibinfo {author} {\bibfnamefont {L.~W.}\ \bibnamefont
  {Tutt}}\ and\ \bibinfo {author} {\bibfnamefont {A.}\ \bibnamefont {Kost}},\
  }\bibfield  {title} {\enquote {\bibinfo {title} {{Optical limiting
  performance of C60 and C70 solutions}},}\ }\href {\doibase 10.1038/356225a0}
  {\bibfield  {journal} {\bibinfo  {journal} {Nature}\ }\textbf {\bibinfo
  {volume} {356}},\ \bibinfo {pages} {225--226} (\bibinfo {year}
  {1992})}\BibitemShut {NoStop}%
\bibitem [{\citenamefont {Belousova}\ \emph {et~al.}(2014)\citenamefont
  {Belousova}, \citenamefont {Videnichev}, \citenamefont {Volynkin},
  \citenamefont {Evstropiev}, \citenamefont {Kislyakov}, \citenamefont
  {Murav'ova},\ and\ \citenamefont {Rakov}}]{Belousova2014}%
  \BibitemOpen
  \bibfield  {author} {\bibinfo {author} {\bibfnamefont {I.~M.}\ \bibnamefont
  {Belousova}}\ \emph {et~al.},\ } 
  \bibfield  {title} {\enquote {\bibinfo {title} {{Nonlinear
  optical limiters of pulsed laser radiation based on carbon-containing
  nanostructures in viscous and solid matrices}},}\ }\href {\doibase
  10.1002/pat.3343} {\bibfield  {journal} {\bibinfo  {journal} {Polym. Adv.
  Technol.}\ }\textbf {\bibinfo {volume} {25}},\ \bibinfo {pages} {1008--1013}
  (\bibinfo {year} {2014})}\BibitemShut {NoStop}%
\bibitem [{\citenamefont {Shanshool}\ \emph {et~al.}(2016)\citenamefont
  {Shanshool}, \citenamefont {Yahaya}, \citenamefont {Yunus},\ and\
  \citenamefont {Abdullah}}]{Shanshool2016}%
  \BibitemOpen
  \bibfield  {author} {\bibinfo {author} {\bibfnamefont {H.~M.}\
  \bibnamefont {Shanshool}}, \bibinfo {author} {\bibfnamefont {M.}\
  \bibnamefont {Yahaya}}, \bibinfo {author} {\bibfnamefont {W.~M.~M.}\
  \bibnamefont {Yunus}}, \ and\ \bibinfo {author} {\bibfnamefont
  {I.~Y.}\ \bibnamefont {Abdullah}},\ }\bibfield  {title} {\enquote
  {\bibinfo {title} {{Influence of polymer matrix on nonlinear optical
  properties and optical limiting threshold of polymer-ZnO nanocomposites}},}\
  }\href {\doibase 10.1007/s10854-016-5001-8} {\bibfield  {journal} {\bibinfo
  {journal} {J. Mater. Sci. Mater. Electron.}\ }\textbf {\bibinfo {volume}
  {27}},\ \bibinfo {pages} {9503--9513} (\bibinfo {year} {2016})}\BibitemShut
  {NoStop}%
\bibitem [{\citenamefont {Mikheev}\ \emph {et~al.}(2016)\citenamefont
  {Mikheev}, \citenamefont {Krivenkov}, \citenamefont {Mikheev}, \citenamefont
  {Okotrub},\ and\ \citenamefont {Mogileva}}]{Mikheev2016}%
  \BibitemOpen
  \bibfield  {author} {\bibinfo {author} {\bibfnamefont {G~M}\ \bibnamefont
  {Mikheev}}, \bibinfo {author} {\bibfnamefont {R.~Yu.}\ \bibnamefont
  {Krivenkov}}, \bibinfo {author} {\bibfnamefont {K.~G.}\ \bibnamefont
  {Mikheev}}, \bibinfo {author} {\bibfnamefont {A.~V.}\ \bibnamefont {Okotrub}},
  \ and\ \bibinfo {author} {\bibfnamefont {T.~N.}\ \bibnamefont {Mogileva}},\
  }\bibfield  {title} {\enquote {\bibinfo {title} {{Z-scanning under
  monochromatic laser pumping: a study of saturatable absorption in a
  suspension of multiwalled carbon nanotubes}},}\ }\href {\doibase
  10.1070/QEL16026} {\bibfield  {journal} {\bibinfo  {journal} {Quantum
  Electron.}\ }\textbf {\bibinfo {volume} {46}},\ \bibinfo {pages} {719--725}
  (\bibinfo {year} {2016})}\BibitemShut {NoStop}%
\bibitem [{\citenamefont {Lim}\ \emph {et~al.}(2011)\citenamefont {Lim},
  \citenamefont {Chen}, \citenamefont {Clark}, \citenamefont {Goh},
  \citenamefont {Ng}, \citenamefont {Tan}, \citenamefont {Friend},
  \citenamefont {Ho},\ and\ \citenamefont {Chua}}]{Lim2011}%
  \BibitemOpen
  \bibfield  {author} {\bibinfo {author} {\bibfnamefont {Geok-Kieng}\
  \bibnamefont {Lim}}\ \emph {et~al.},\ }\bibfield  {title}
  {\enquote {\bibinfo {title} {{Giant broadband nonlinear optical absorption
  response in dispersed graphene single sheets}},}\ }\href {\doibase
  10.1038/nphoton.2011.177} {\bibfield  {journal} {\bibinfo  {journal} {Nature
  Photon.}\ }\textbf {\bibinfo {volume} {5}},\ \bibinfo {pages} {554--560}
  (\bibinfo {year} {2011})}\BibitemShut {NoStop}%
\bibitem [{\citenamefont {Tolbin}\ \emph {et~al.}(2016)\citenamefont {Tolbin},
  \citenamefont {Savelyev}, \citenamefont {Gerasimenko},\ and\ \citenamefont
  {Tomilova}}]{Tolbin2016}%
  \BibitemOpen
  \bibfield  {author} {\bibinfo {author} {\bibfnamefont {A.~Yu.}\
  \bibnamefont {Tolbin}}, \bibinfo {author} {\bibfnamefont {M.~S.}\
  \bibnamefont {Savelyev}}, \bibinfo {author} {\bibfnamefont {A.~Yu.}\
  \bibnamefont {Gerasimenko}}, \ and\ \bibinfo {author} {\bibfnamefont
  {L.~G.}\ \bibnamefont {Tomilova}},\ }\bibfield  {title} {\enquote
  {\bibinfo {title} {{High-performance optical limiters based on stable
  phthalocyanine J-type dimers}},}\ }\href {\doibase
  10.1016/j.cplett.2016.06.051} {\bibfield  {journal} {\bibinfo  {journal}
  {Chem. Phys. Lett.}\ }\textbf {\bibinfo {volume} {661}},\ \bibinfo {pages}
  {269--273} (\bibinfo {year} {2016})}\BibitemShut {NoStop}%
\bibitem [{\citenamefont {Ryzhov}\ and\ \citenamefont
  {Belousova}(2015)}]{Ryzhov2015}%
  \BibitemOpen
  \bibfield  {author} {\bibinfo {author} {\bibfnamefont {A.~A.}\ \bibnamefont
  {Ryzhov}}\ and\ \bibinfo {author} {\bibfnamefont {I.~M.}\ \bibnamefont
  {Belousova}},\ }\bibfield  {title} {\enquote {\bibinfo {title} {{Optical
  Limiting Characteristics of Fabry–Perot Microresonators at Third-order
  Nonlinear Absorption and Refraction of the Intracavity Medium}},}\ }in\ \href
  {\doibase 10.5220/0005404301400143} {\emph {\bibinfo {booktitle} {Proc. 3rd
  Int. Conf. Photonics, Opt. Laser Technol.}}}\ (\bibinfo {year} {2015})\ pp.\
  \bibinfo {pages} {140--143}\BibitemShut {NoStop}%
\bibitem [{\citenamefont {Siegman}(1962)}]{Siegman1962}%
  \BibitemOpen
  \bibfield  {author} {\bibinfo {author} {\bibfnamefont {A.~E.}\ \bibnamefont
  {Siegman}},\ }\bibfield  {title} {\enquote {\bibinfo {title} {{Nonlinear
  Optical Effects: An Optical Power Limiter}},}\ }\href {\doibase
  10.1364/AO.1.000739} {\bibfield  {journal} {\bibinfo  {journal} {Appl. Opt.}\
  }\textbf {\bibinfo {volume} {1}},\ \bibinfo {pages} {739} (\bibinfo {year}
  {1962})}\BibitemShut {NoStop}%
\bibitem [{\citenamefont {Ryzhov}\ \emph {et~al.}(2014)\citenamefont {Ryzhov},
  \citenamefont {Belousova}, \citenamefont {Wang}, \citenamefont {Qi},\ and\
  \citenamefont {Wang}}]{Ryzhov2014}%
  \BibitemOpen
  \bibfield  {author} {\bibinfo {author} {\bibfnamefont {A.~A.}\ \bibnamefont
  {Ryzhov}}, \bibinfo {author} {\bibfnamefont {I.~M.}\ \bibnamefont
  {Belousova}}, \bibinfo {author} {\bibfnamefont {Ya.}~\bibnamefont {Wang}},
  \bibinfo {author} {\bibfnamefont {H.}~\bibnamefont {Qi}}, \ and\ \bibinfo
  {author} {\bibfnamefont {J.}~\bibnamefont {Wang}},\ }\bibfield  {title}
  {\enquote {\bibinfo {title} {{Optical limiting properties of a nonlinear
  multilayer Fabry–Perot resonator containing niobium pentoxide as nonlinear
  medium}},}\ }\href {\doibase 10.1364/OL.39.004847} {\bibfield  {journal}
  {\bibinfo  {journal} {Opt. Lett.}\ }\textbf {\bibinfo {volume} {39}},\
  \bibinfo {pages} {4847--50} (\bibinfo {year} {2014})}\BibitemShut {NoStop}%
\bibitem [{\citenamefont {Valligatla}\ \emph {et~al.}(2015)\citenamefont
  {Valligatla}, \citenamefont {Chiasera}, \citenamefont {Varas}, \citenamefont
  {Das}, \citenamefont {{Shivakiran Bhaktha}}, \citenamefont {{\L}ukowiak},
  \citenamefont {Scotognella}, \citenamefont {{Narayana Rao}}, \citenamefont
  {Ramponi}, \citenamefont {Righini},\ and\ \citenamefont
  {Ferrari}}]{Valligatla2015}%
  \BibitemOpen
  \bibfield  {author} {\bibinfo {author} {\bibfnamefont {S.}\
  \bibnamefont {Valligatla}}\ \emph {et~al.},\ }\bibfield  {title}
  {\enquote {\bibinfo {title} {{Optical field enhanced nonlinear absorption and
  optical limiting properties of 1-D dielectric photonic crystal with ZnO
  defect}},}\ }\href {\doibase 10.1016/j.optmat.2015.10.032} {\bibfield
  {journal} {\bibinfo  {journal} {Opt. Mater. (Amst).}\ }\textbf {\bibinfo
  {volume} {50}},\ \bibinfo {pages} {229--233} (\bibinfo {year}
  {2015})}\BibitemShut {NoStop}%
\bibitem [{\citenamefont {Dong}\ \emph {et~al.}(2015)\citenamefont {Dong},
  \citenamefont {Li}, \citenamefont {Feng}, \citenamefont {Zhang},
  \citenamefont {Zhang}, \citenamefont {Chang}, \citenamefont {Fan},
  \citenamefont {Zhang},\ and\ \citenamefont {Wang}}]{Dong2015}%
  \BibitemOpen
  \bibfield  {author} {\bibinfo {author} {\bibfnamefont {N.}\
  \bibnamefont {Dong}}\ \emph {et~al.},\ }\bibfield  {title} {\enquote
  {\bibinfo {title} {{Optical Limiting and Theoretical Modelling of Layered
  Transition Metal Dichalcogenide Nanosheets}},}\ }\href {\doibase
  10.1038/srep14646} {\bibfield  {journal} {\bibinfo  {journal} {Sci. Rep.}\
  }\textbf {\bibinfo {volume} {5}},\ \bibinfo {pages} {14646} (\bibinfo {year}
  {2015})}\BibitemShut {NoStop}%
\bibitem [{\citenamefont {Vella}\ \emph {et~al.}(2016)\citenamefont {Vella},
  \citenamefont {Goldsmith}, \citenamefont {Browning}, \citenamefont
  {Limberopoulos}, \citenamefont {Vitebskiy}, \citenamefont {Makri},\ and\
  \citenamefont {Kottos}}]{Vella2016}%
  \BibitemOpen
  \bibfield  {author} {\bibinfo {author} {\bibfnamefont {J.~H.}\
  \bibnamefont {Vella}}\ \emph {et~al.},\ }\bibfield  {title} {\enquote {\bibinfo {title}
  {{Experimental Realization of a Reflective Optical Limiter}},}\ }\href
  {\doibase 10.1103/PhysRevApplied.5.064010} {\bibfield  {journal} {\bibinfo
  {journal} {Phys. Rev. Appl.}\ }\textbf {\bibinfo {volume} {5}},\ \bibinfo
  {pages} {1--7} (\bibinfo {year} {2016})}\BibitemShut {NoStop}%
\bibitem [{\citenamefont {Adachi}(1985)}]{Adachi1985}%
  \BibitemOpen
  \bibfield  {author} {\bibinfo {author} {\bibfnamefont {S.}\ \bibnamefont
  {Adachi}},\ }\bibfield  {title} {\enquote {\bibinfo {title} {{GaAs, AlAs, and
  AlGaAs Material parameters for use in research and device applications}},}\
  }\href {\doibase 10.1063/1.336070} {\bibfield  {journal} {\bibinfo  {journal}
  {J. Appl. Phys.}\ }\textbf {\bibinfo {volume} {58}},\ \bibinfo {pages} {R1}
  (\bibinfo {year} {1985})}\BibitemShut {NoStop}%
\bibitem [{\citenamefont {Deri}\ and\ \citenamefont
  {Emanuel}(1995)}]{Deri1995}%
  \BibitemOpen
  \bibfield  {author} {\bibinfo {author} {\bibfnamefont {R.~J.}\ \bibnamefont
  {Deri}}\ and\ \bibinfo {author} {\bibfnamefont {M.~A.}\ \bibnamefont
  {Emanuel}},\ }\bibfield  {title} {\enquote {\bibinfo {title} {{Consistent
  formula for the refractive index of AlGaAs below the band edge}},}\ }\href
  {\doibase 10.1063/1.359434} {\bibfield  {journal} {\bibinfo  {journal} {J.
  Appl. Phys.}\ }\textbf {\bibinfo {volume} {77}},\ \bibinfo {pages} {4667}
  (\bibinfo {year} {1995})}\BibitemShut {NoStop}%
\bibitem [{\citenamefont {Said}\ \emph {et~al.}(1992)\citenamefont {Said},
  \citenamefont {Sheik-Bahae}, \citenamefont {Hagan}, \citenamefont {Wei},
  \citenamefont {Wang}, \citenamefont {Young},\ and\ \citenamefont
  {Stryland}}]{Said1992}%
  \BibitemOpen
  \bibfield  {author} {\bibinfo {author} {\bibfnamefont {A.~A.}\ \bibnamefont
  {Said}}\ \emph {et~al.},\ }\bibfield  {title} {\enquote
  {\bibinfo {title} {{Determination of bound-electronic and free-carrier
  nonlinearities in ZnSe, GaAs, CdTe, and ZnTe}},}\ }\href {\doibase
  10.1364/JOSAB.9.000405} {\bibfield  {journal} {\bibinfo  {journal} {J. Opt.
  Soc. Am. B}\ }\textbf {\bibinfo {volume} {9}},\ \bibinfo {pages} {405}
  (\bibinfo {year} {1992})}\BibitemShut {NoStop}%
\bibitem [{\citenamefont {Gonzalez}\ \emph {et~al.}(2009)\citenamefont
  {Gonzalez}, \citenamefont {Murray}, \citenamefont {Krishnamurthy},\ and\
  \citenamefont {Guha}}]{Gonzalez2009}%
  \BibitemOpen
  \bibfield  {author} {\bibinfo {author} {\bibfnamefont {L.~P.}\
  \bibnamefont {Gonzalez}}, \bibinfo {author} {\bibfnamefont {J.~M.}\
  \bibnamefont {Murray}}, \bibinfo {author} {\bibfnamefont {S.}\
  \bibnamefont {Krishnamurthy}}, \ and\ \bibinfo {author} {\bibfnamefont
  {S.}\ \bibnamefont {Guha}},\ }\bibfield  {title} {\enquote {\bibinfo
  {title} {{Wavelength dependence of two photon and free carrier absorptions in
  InP}},}\ }\href {\doibase 10.1364/OE.17.008741} {\bibfield  {journal}
  {\bibinfo  {journal} {Opt. Express}\ }\textbf {\bibinfo {volume} {17}},\
  \bibinfo {pages} {8741} (\bibinfo {year} {2009})}\BibitemShut {NoStop}%
\bibitem [{\citenamefont {Boggess}\ \emph {et~al.}(1985)\citenamefont
  {Boggess}, \citenamefont {Smirl}, \citenamefont {Moss}, \citenamefont
  {Boyd},\ and\ \citenamefont {{Van Stryland}}}]{Boggess1985}%
  \BibitemOpen
  \bibfield  {author} {\bibinfo {author} {\bibfnamefont {T.}~\bibnamefont
  {Boggess}}, \bibinfo {author} {\bibfnamefont {A.}~\bibnamefont {Smirl}},
  \bibinfo {author} {\bibfnamefont {S.}~\bibnamefont {Moss}}, \bibinfo {author}
  {\bibfnamefont {I.}~\bibnamefont {Boyd}}, \ and\ \bibinfo {author}
  {\bibfnamefont {E.}~\bibnamefont {{Van Stryland}}},\ }\bibfield  {title}
  {\enquote {\bibinfo {title} {{Optical limiting in GaAs}},}\ }\href {\doibase
  10.1109/JQE.1985.1072688} {\bibfield  {journal} {\bibinfo  {journal} {IEEE J.
  Quantum Electron.}\ }\textbf {\bibinfo {volume} {21}},\ \bibinfo {pages}
  {488--494} (\bibinfo {year} {1985})}\BibitemShut {NoStop}%
\bibitem [{\citenamefont {Andrews}\ \emph {et~al.}(2016)\citenamefont
  {Andrews}, \citenamefont {Smotzer}, \citenamefont {Latronica},\ and\
  \citenamefont {Crescimanno}}]{Andrews2016}%
  \BibitemOpen
  \bibfield  {author} {\bibinfo {author} {\bibfnamefont {J.~H.}\
  \bibnamefont {Andrews}}, \bibinfo {author} {\bibfnamefont {M.}\
  \bibnamefont {Smotzer}}, \bibinfo {author} {\bibfnamefont {B.}\
  \bibnamefont {Latronica}}, \ and\ \bibinfo {author} {\bibfnamefont {M.}\
  \bibnamefont {Crescimanno}},\ }\bibfield  {title} {\enquote {\bibinfo {title}
  {{Linear distributed Bragg cavity effects on optical limiting in two- and
  three-level media}},}\ }\href {\doibase 10.1364/JOSAB.33.00E102} {\bibfield
  {journal} {\bibinfo  {journal} {J. Opt. Soc. Am. B}\ }\textbf {\bibinfo
  {volume} {33}},\ \bibinfo {pages} {E102} (\bibinfo {year}
  {2016})}\BibitemShut {NoStop}%
\end{thebibliography}

\end{document}